\documentclass[a4paper,USenglish,autoref]{lipics-v2019}

\hypersetup{
  colorlinks=true,
  pdfstartview=FitV,
  linkcolor=blue,
  citecolor=blue,
  urlcolor=blue
}

\usepackage{stmaryrd}
\usepackage{thm-restate}
\usepackage{mathtools}
\usepackage{tikz}
\usetikzlibrary{automata,arrows}
\usetikzlibrary{positioning}

\usepackage[ruled,linesnumbered,vlined]{algorithm2e}
\DontPrintSemicolon
\SetNlSty{bfseries}{\color{black}}{}
\SetKwProg{Function}{Function}{:}{}
\usepackage{colonequals}
\newcommand{\assign}{\colonequals}

\addto\extrasUSenglish{

}

\newcommand\booleans{\mathbb B}
\newcommand\integers{\mathbb Z}
\newcommand\naturals{\mathbb N}

\newcommand\A{\mathcal A}
\newcommand\M{\mathcal M}
\renewcommand\P{\mathcal P}
\newcommand\jump[1]{\xrightarrow{#1}}

\newcommand\oword{$\omega$-word}

\newcommand\semantics[1]{\llbracket #1 \rrbracket}
\newcommand{\set}[1]{\{#1\}}
\newcommand{\tup}[1]{\langle#1\rangle}

\newcommand\PP{\mathbb P}
\newcommand\EE{\mathbb E}
\newcommand{\asto}{\xrightarrow{\text{a.s.}}}
\newcommand{\as}{\qquad \text{as }}
\newcommand{\iid}{i.i.d.}

\newcommand{\Markov}{\mathit{Markov}}

\newcommand{\mcm}{m_q}
\newcommand{\mcV}{V_q^n}
\newcommand{\mcT}[1]{{T_q^n}^{(#1)}}
\newcommand{\mcS}[1]{{S_q^n}^{(#1)}}
\newcommand{\mcSS}{\overline{\mathit{SS}}_q^n}
\newcommand{\mcTS}{\overline{\mathit{TS}}_q^n}

\newcommand{\offset}{s}
\newcommand{\freq}{f}

\newcommand{\infixCnt}{\gamma}

\newcommand{\always}{\Box}
\newcommand{\eventually}{\Diamond}

\let\epsilon\varepsilon

\DeclarePairedDelimiter\ceil{\lceil}{\rceil}
\DeclarePairedDelimiter\floor{\lfloor}{\rfloor}

\newcommand\mode{\mathsf{mode}}
\newcommand\median{\mathsf{median}}

\newcommand\slt{\prec}
\newcommand\sgt{\succ}
\newcommand\sle{\preccurlyeq}
\newcommand\sge{\succcurlyeq}

\newcommand{\twopartdef}[4]
{
  \left\{
    \begin{array}{ll}
      #1 & \mbox{if } #2 \\
      #3 & \mbox{if } #4
    \end{array}
  \right.
}

\title{Monitoring Event Frequencies}

\author{Thomas Ferr\`{e}re}{IST Austria, Klosterneuburg, Austria}{}{https://orcid.org/0000-0001-5199-3143}{}
\author{Thomas A. Henzinger}{IST Austria, Klosterneuburg, Austria}{}{https://orcid.org/0000-0002-2985-7724}{}
\author{Bernhard Kragl}{IST Austria, Klosterneuburg, Austria}{}{https://orcid.org/0000-0001-7745-9117}{}

\authorrunning{T. Ferr\`{e}re, T.\,A. Henzinger, and B. Kragl}

\Copyright{Thomas Ferr\`{e}re, Thomas A. Henzinger, and Bernhard Kragl}



\relatedversion{This is an extended version of a paper presented at the 28th EACSL Annual Conference on Computer Science Logic (CSL 2020), which provides missing proofs in the appendix. The conference version is available at \url{https://doi.org/10.4230/LIPIcs.CSL.2020.20}.}

\funding{This research was supported in part by the Austrian Science Fund (FWF) under grants S11402-N23 (RiSE/SHiNE) and Z211-N23 (Wittgenstein Award).}

\acknowledgements{We thank Jan Maas for showing us a simple proof of convergence of the mode in the \iid\ case, and Zbigniew S. Szewczak for pointing out to us the use of Karamata's Tauberian theorem in connection with ergodic theory~\cite{Szewczak08}.}

\nolinenumbers

\hideLIPIcs

\EventEditors{Maribel Fern\'{a}ndez and Anca Muscholl}
\EventNoEds{2}
\EventLongTitle{28th EACSL Annual Conference on Computer Science Logic (CSL 2020)}
\EventShortTitle{CSL 2020}
\EventAcronym{CSL}
\EventYear{2020}
\EventDate{January 13--16, 2020}
\EventLocation{Barcelona, Spain}
\EventLogo{}
\SeriesVolume{152}
\ArticleNo{20}

\begin{document}

\maketitle

\begin{abstract}
The monitoring of event frequencies can be used to recognize behavioral anomalies, to identify trends, and to deduce or discard hypotheses about the underlying system.
For example, the performance of a web server may be monitored based on the ratio of the total count of requests from the least and most active clients.
Exact frequency monitoring, however, can be prohibitively expensive; in the above example it would require as many counters as there are clients.
In this paper, we propose the efficient probabilistic monitoring of common frequency properties, including the mode (i.e., the most common event) and the median of an event sequence.
We define a logic to express composite frequency properties as a combination of atomic frequency properties.
Our main contribution is an algorithm that, under suitable probabilistic assumptions, can be used to monitor these important frequency properties with four counters, independent of the number of different events.
Our algorithm samples longer and longer subwords of an infinite event sequence.
We prove the almost-sure convergence of our algorithm by generalizing ergodic theory from increasing-length prefixes to increasing-length subwords of an infinite sequence.
A similar algorithm could be used to learn a connected Markov chain of a given structure from observing its outputs, to arbitrary precision, for a given confidence.


\end{abstract}

\section{Introduction}\label{sec:introduction}
The safety and security of computerized systems are ensured
by a chain of methods that use logic and formal semantics to assert and check the correct operation of a system, real or simulated.
Runtime monitoring~\cite{RV-survery} happens at the end of this chain and complements rigorous design and verification practices to catch malfunctions as they occur in a live system.
In addition to critical functional aspects, softer performance metrics also need to be monitored to ensure a suitable quality of service.
Monitoring system properties takes place in
parallel with the execution of the system itself.
A dedicated component, called monitor, observes the system behavior as input and generate a verdict about the system behavior as output.
Due to reactivity considerations, the monitor is often required to
perform its observations in real time, and not being the
main computational artifact, should consume limited resources.

In this paper we propose a new class of quantitative properties based on event frequencies, called \emph{frequency properties}, and study their monitoring problem.
In particular, we define a logic to express composite frequency properties as linear and Boolean combinations of atomic frequency properties.
While all such frequency properties are theoretically monitorable using counter registers, there are, in general, no efficient monitoring algorithms in the case of large or infinite input alphabets.
As a motivating example we use the \emph{mode} of a sequence over a finite alphabet~$\Sigma$.
By definition, $a \in \Sigma$ is the mode of an $\omega$-word $w$ if there exists a length
$n$ such that each prefix of $w$ longer than $n$ contains more occurrences of $a$'s
than occurrences of any other letter $b\in\Sigma$.
This frequency property can be monitored using a separate counter for every event in~$\Sigma$.
However, the alphabet $\Sigma$ is typically too large for this to be practical.\footnote{Consider the IPv4 protocol alphabet with its 4,294,967,296 letters (addresses) and
the UTF-8 encoding alphabet with its 1,112,064 letters (code points).}
We show that there is no shortcut to monitor the mode exactly
and in real time:
in general $|\Sigma|$ counters are needed for this task.

However, we are not always interested in monitoring exactly and in real
time the mode after every new event, and sometimes wish to
estimate what the mode is expected to be in the future.
Perhaps surprisingly, we can then do much better.
Let us assume that the past, finite, observed behavior of an
event sequence is representative of the future, infinite, unknown behavior.
This is the case for stochastic systems, for instance if the observation
sequence is generated by a Markov chain.
We move from the \emph{real-time monitoring} problem, asking
to compute or approximate, in real time, the value of a frequency
property for each observed prefix, to the \emph{limit monitoring} problem,
asking to estimate the future limit value of the frequency property,
if it exists.
In particular, for the mode of a connected Markov chain, the longer we
observe a behavior, the higher our confidence in predicting its mode.
While every real-time monitor can be used as limit monitor, there can
be limit monitors that use dramatically fewer resources.

We present a simple, memory-efficient strategy to limit monitor frequency properties of random $\omega$-words.
In particular, our mode monitor uses four counters only.
Two of the counters keep track of the number of occurrences of two letters at
a time.
The first letter is the current mode prediction, say~$a$.
The second letter is the mode replacement candidate, say~$b$.
We count the number of $a$'s and $b$'s over a given subword, until a certain
number of events, say 10, has been processed.
The most frequent letter out of $a$ and $b$ in this 10-letter subword, say~$a$, wins the round
and becomes the new mode prediction.
The other letter loses the round and is replaced by a letter sampled at
random, say~$c$.
In the next round the subword length will be increased, say to~11, and
$a$ will compete against $c$ over the next subword.
We reuse two counters for the two letters, and the other two counters to
keep track of the current subword length and to stop counting when that
length is reached.
By repeating the process we get increasingly higher confidence that $a$ is
indeed the mode.
Even if by random perturbation the mode $a$ of the generating Markov chain
was no longer the current prediction, it would eventually get sampled again
and statistically reappear, and eventually remain, as the prediction.

The algorithm of our mode monitor easily transfers to an efficient monitor for the median.
Indeed, we also show that our results generalize to any property expressible as Boolean combination of linear inequalities over frequencies of events.
An application of our algorithmic ideas is to learn the transition probabilities of a connected Markov chain of known structure through the observation of subword frequencies.

The main result of this paper is that, assuming the monitored system
is a connected Markov chain, our monitoring algorithm converges
almost surely.
The proof of this fact calls for a new ergodic theory based on subwords as opposed to prefixes.
This theory uses as its main building block a variant of the law of large
numbers over so-called triangular random arrays of the form
$X_{1,1},X_{2,1},X_{2,2},X_{3,1},\dots$ and hinges on deep results from
matrix theory.
The correctness of the algorithm can also be understood, in a weaker form, by showing
convergence in probability of its output.
Assuming that the Markov chain starts in a stationary distribution,
the probability of a given word $u$ occurring as
subword of an $\omega$-word $w$ at position $i$ is independent of~$i$.
As a result, when the value of a function over prefixes converges
probabilistically, then the same limit is reached probabilistically over
arbitrary subwords.

In short, the main conceptual and technical contributions of this paper are the following:
\begin{enumerate}
\item We show that precise real-time monitoring is inherently resource-intensive (\autoref{sec:precise}).
\item We propose the novel setting of limit monitoring (\autoref{sec:monitoring}).
\item We provide a generic scheme for efficient limit monitoring (\autoref{sec:efficient}) and instantiate it to specialized monitoring algorithms for the mode (\autoref{subsec:mode}) and median (\autoref{subsec:median}).
\item We define a logic for composite frequency properties which combines atomic frequency properties such that each formula of the logic can be limit monitored efficiently (\autoref{sec:general}).
\item We develop a new ergodic theory for connected Markov chains (\autoref{subsec:ergodic}) to prove our monitoring algorithms correct.
\end{enumerate}

\subsection{Related Work}
In the area of formal verification, probabilistic model checking~\cite{Kwiatkowska07,KwiatkowskaNP2018} and quantitative verification~\cite{Henzinger13} are concerned with the white-box static analysis of a probabilistic system.
Statistical model checking~\cite{AghaP18} tries to learn the probabilistic structure of a system by sampling \emph{many} executions, and thus also applies to black-box systems.
These are in contrast to our monitoring setting where a \emph{single} execution of a black-box system is dynamically observed during execution.
Our work belongs specifically to the field of runtime verification~\cite{RV-survery}, which is concerned with the evaluation of temporal properties over program traces.
While much of the research in this domain assumes finite-state monitors, in this work we study an infinite-state problem based on the model of counter monitors. 
The expressiveness of different register machines and resource trade-offs for monitoring safety properties involving counters and arithmetic registers is studied in~\cite{FerrereHS18}.
Another infinite-state model for monitoring is that of quantified event automata~\cite{BarringerFHRR12}, which combine finite automata specifications with first-order quantification.
Other quantitative automata machines are surveyed in~\cite{ChatterjeeHO16}.

\looseness=-1
The computation of aggregates over an ongoing system execution in real time was considered in various areas of computer science.
Stream expressions~\cite{DAngeloSSRFSMM05,RTLola-arxiv} and quantitative regular expressions~\cite{AlurFR16} provide frameworks for the specification of transducers over data streams.
The work on runtime verification and stream processing can be seen as solving real-time monitoring problems, and very rarely assumes a probabilistic model.
A notable exception can be found in~\cite{SammapunLS05}, who propose to use hypothesis testing to provide an interval of confidence on the monitor outcome when evaluating some probabilistic property.
In the vast literature from runtime verification to online algorithms, the problem of limit monitoring as defined, solved, and applied in this paper was, to the best of our knowledge, not studied before.

It is well-known that certain common statistical indicators can be computed in real time.
For example, the average can be computed by simply maintaining the sum and sample size.
Perhaps more surprisingly, the variance and covariance of a sequence can also be computed in one pass through classical online algorithms~\cite{Welford62}.
However, other indicators, like the median, are hard or impossible to compute in real time.
Offline algorithms for the median include selection algorithms (e.g., quickselect~\cite{Hoare61a}) with $O(n)$ run time (versus $O(n \log n)$ for sorting), median of medians~\cite{BlumFPRT73} (which is approximate), and the randomized algorithm of Mitzenmacher \& Upfal~\cite{MitzenmacherU05}.
The best known online algorithm uses two heaps to store the lower and higher half of values (i.e., all samples have to be stored), with an amortized cost of $O(\log n)$ per input.
To the best of our knowledge, no real-time algorithm to compute the median exactly was proposed in the literature.

Statistical properties of subword frequencies in Markov chains are studied in~\cite{Booth68}.
In Markov chain theory, the existence, uniqueness, and convergence results for stationary distributions are among the most fundamental results~\cite{Norris98}.
The rate of convergence towards a stationary distribution is called mixing time~\cite{MixingTimes}.
In general, the mixing time is controlled by the spectral gap of the transition matrix, with precise results only know for particular random processes, like card shuffling.
These result do not lead to bounds on the convergence rate of frequencies of events in labeled Markov chains.

An indirect (and somewhat degenerate) approach to monitoring would be to first learn the monitored system, and then perform offline verification on the learned model.
Learning probabilistic generators was studied in the setting of automata learning~\cite{RonThesis}, but requires more powerful oracle queries like membership and equivalence.
Rudich showed that the structure and transition probabilities of a Markov chain can, in principle, be learned from a single input sequence~\cite{Rudich85}.
However, the algorithm is impractical as it essentially enumerates all possible structures.



\section{Definitions}\label{sec:definitions}
Let $\Sigma$ be a finite alphabet of events.
Given a finite or infinite word or $\omega$-word $w \in \Sigma^* \cup \Sigma^\omega$ and a position $i$, $1 \le i \le |w|$, we denote by $w_i$ its $i$'th value.
Given a pair of positions $i$ and $j$, $i \le j$, we denote by $w_{i..j}$ the \emph{infix} of $w$ from $i$ to $j$, such that $|w_{i..j}| = j-i+1$ and $(w_{i..j})_k = w_{i+k-1}$ for all $1 \le k \le j-i+1$.
We denote by $w_{..i} = w_{1..i}$ the \emph{prefix} of $w$ of length $i$.
For any word $w \in \Sigma^*$ and letter $a \in \Sigma$ we write $|w|_a$ for the number of occurrences of $a$ in $w$.

\subsection{Sequential Statistics}

We define a statistic to be any function that outputs an indicator for a given input word.

\begin{definition}[Statistic]
Let $\Sigma$ be a finite alphabet and $\Lambda$ be an output domain.
A \emph{statistic} is a function $\mu : \Sigma^* \to \Lambda$.
\end{definition}

In this paper we focus on statistics that are based on the frequency, or number of occurrence, of events.
Two typical examples are the \emph{mode}, i.e. 
the most frequent event, and the \emph{median}, i.e., the value separating as evenly as possible the upper half from the lower half of a data sample.

\begin{example}[Mode]\label{ex:mode}
We say that $a \in \Sigma$ is the \emph{mode} of $w$ when $|w|_a > |w|_\sigma$ for all $\sigma \in \Sigma \setminus \set{a}$.
We denote by $\mode : \Sigma^* \to \Sigma \uplus \set{\bot}$ the statistic that maps a word to its mode if it exists, or to $\bot$ otherwise.
\end{example}

\begin{example}[Median]\label{ex:median}
Let $\Sigma$ be ordered by $\slt$.
We say that $a \in \Sigma$ is the \emph{median} of $w$ when $\sum_{\sigma \sgt a} |w|_\sigma < \sum_{\sigma \sle a} |w|_\sigma$ and $\sum_{\sigma \slt a} |w|_\sigma < \sum_{\sigma \sge a} |w|_{\sigma}$.
We denote by $\median : \Sigma^* \to \Sigma \uplus \set{\bot}$ the statistic that maps a word to its median if it exists, or to $\bot$ otherwise.
\end{example}

An example of a statistic that takes into account the order of events in a word is the most frequent event that occurs right after some dedicated event.

\subsection{Counter Monitors}

The task of a monitor is to compute a statistic in real time.
We define a variant of monitor machines that allows us to classify a monitor based on the amount of resources it uses.
We adapt the definition of counter monitors set in~\cite{FerrereHS18} to our setting of monitoring frequencies.

Let $X$ be a set of integer variables, called \emph{registers} or \emph{counters}.
Registers can be read and written according to relations and functions in the signature $S=\tup{0,+1,\le}$ as follows:
\begin{itemize}
\item A \emph{test} is a conjunction of atomic formulas over $S$ and their negation;
\item An \emph{update} is a mapping from variables to terms over $S$.
\end{itemize}
The set of tests and updates over $X$ are denoted $\Phi(X)$ and $\Gamma(X)$, respectively.

\begin{definition}[Counter Monitor]
A \emph{counter monitor} is a tuple $\A = (\Sigma,\Lambda,X,Q,\lambda,s,\Delta)$, where $\Sigma$ is an input alphabet, $\Lambda$ is an output alphabet, $X$ is a set of registers, $Q$ is a set of control locations, $\lambda : Q \times \naturals^X \to \Lambda$ is an output function, $s \in Q$ is the initial location, and $\Delta \subseteq Q \times \Sigma \times \Phi(X) \times \Gamma(X) \times Q$ is a transition relation such that for every location $q \in Q$, event $\sigma \in \Sigma$, and valuation $v : X \to \naturals$ there exists a unique edge $(q,\sigma,\phi,\gamma,q') \in \Delta$ such that $v \models \phi$ is satisfied.
The sets $\Sigma, X, Q, \Delta$ are assumed to be finite.
\end{definition}

A \emph{run} of the monitor $\A$ over a word $w \in \Sigma^* \cup \Sigma^\omega$ is a sequence of transitions $(q_1,v_1) \jump{w_1} (q_2,v_2) \jump{w_2} \hdots$ labeled by $w$ such that $q_1 = s$ and $v_1(x) = 0$ for all $x \in X$. Here we write $(q,v) \jump{\sigma} (q',v')$ when there exists an edge $(q,\sigma,\phi,\gamma,q') \in \Delta$ such that $v \models \phi$ and $v'(x) = v(\gamma(x))$ for all $x \in X$.
There exists exactly one run of a given counter monitor $\A$ over a given word $w$.

\begin{definition}[Monitor Semantics]
  Every counter monitor $\A$ computes a statistic $\semantics\A : \Sigma^* \to \Lambda$, such that $\semantics\A(w) = \lambda(q,v)$ for $(q,v)$ the final state in the run of $\A$ over $w \in \Sigma^*$.
\end{definition}

We remark that the term ``counter machine'' has various different meanings in the literature and designates machines with varying computational power.
In our definition we note the use of the constant $0$ which enables resets.
Such resets cannot be simulated in real time. 
On the contrary, arbitrary increments are w.l.o.g., as shown in~\cite{FischerMR68}.

\subsection{Probabilistic Generators}

In this work we model systems as labeled Markov chains, whose executions generate random $\omega$-words.

\begin{definition}[Markov Chain]
A (finite, connected, labeled) \emph{Markov chain} is a tuple $\M = (\Sigma,Q,\lambda,\pi,p)$, where $\Sigma$ is a finite set of events, $Q$ is a finite set of states, $\lambda : Q \to \Sigma$ is a labeling, $\pi$ is an initial-state distribution over $Q$, and $p : Q \times Q \to [0,1]$ is a transition distribution with $\sum_{q' \in Q} p(q,q') = 1$ for all $q \in Q$ and whose set of edges $(q,q')$ such that $p(q,q') > 0$ forms a strongly connected graph.
\end{definition}

In the rest of this paper, even when not explicitly stated, every Markov chain is assumed to be finite and connected.

Let $\M = (\Sigma,Q,\lambda,\pi,p)$ be a Markov chain. A random infinite sequence $(X_i)_{i \ge 1}$ of states is an execution of $\M$,
$\Markov(\M)$ for short,
if (i)~$X_1$ has distribution $\pi$ and (ii)~conditional on $X_i = q$, $X_{i+1}$ has distribution $q' \mapsto p(q,q')$ and is independent of $X_1,\dots,X_{i-1}$.
By extension, a random $\omega$-word $w$ is $\Markov(\M)$ if $w_i = \lambda(X_i)$ for all $i \ge 1$.

We denote by $V_q(k) = \sum_{i=1}^{k} 1_{\set{X_i = q}}$ the \emph{number of visits} to state $q$ within $k$ steps, and by $T_q = \inf\set{i > 1 \mid X_i = q}$ the \emph{first time of visiting} state $q$ (after the initial state).
Then $\mcm = \EE(T_q \mid X_1=q)$ is the \emph{expected return time} to state $q$.
The ergodic theorem for Markov chains states that the long-run proportion of time spent in each state $q$ is the inverse of $m_q$.
Thus we call $\freq_q = \frac 1 \mcm$ the (long-run) \emph{frequency} of $q$.

\begin{theorem}[Ergodic Theorem~\cite{Norris98}]\label{thm:ergodic-classic}
  Let $\M$ be a finite connected Markov chain.
  If $(X_i)_{i \ge 1}$ is $\mathit{Markov}(\M)$ then
  $V_q(n) / n \asto \freq_q$ as $n \to \infty$
  for every state $q$.
\end{theorem}

Now summing the frequencies of all states mapped to a letter $\sigma$ gives the expected frequency of $\sigma$,
$\freq_\sigma = \sum_{\substack{q \in Q \\ \lambda(q)=\sigma}} \freq_q$,
as characterized by the following corollary.

\begin{corollary}\label{cor:ergodic-classic}
  Let $\M$ be a finite connected Markov chain.
  If $w$ is $\mathit{Markov}(\M)$ then
  ${|w_{..n}|_\sigma} / n \asto \freq_\sigma$ as $n \to \infty$
  for every letter $\sigma$.
\end{corollary}


\section{The Limit-Monitoring Problem}\label{sec:monitoring}
We want to monitor the value of a given statistic $\mu : \Sigma^* \to \Lambda$ over the execution of some (probabilistic) process $\P$.
This execution is potentially infinite, forming an \oword\ $w \in \Sigma^\omega$.
In practice, the statistic $\mu$ is often used as an estimator of some parameter $v \in \Lambda$ of process $\P$. 
Such a parameter is always well-defined in the case where $\mu$ converges to $v$ as follows.
\begin{definition}[Convergence]\label{def:convergence}
  A statistic $\mu : \Sigma^* \to \Lambda$ \emph{(almost surely) converges} to a value $v \in \Lambda$ over a random process $\P$, written $\mu(\P) = v$, if $\PP_{w \sim \P}(\lim_{n \to \infty} \mu(w_{..n}) = v) = 1$.
\end{definition}

Computing the value of the statistic $\mu$ over every finite prefix of $w$ can be an objective in itself.
It gives us the most precise estimate of the parameter $v$ when defined.
A monitor fulfilling this requirement is called \emph{real-time}.
Such a monitor is past-oriented, and is concerned with computing accurately the value $\mu(w_{..n})$ of the statistic at step $n$, for all $n$.
\begin{definition}[Real-Time Monitoring]
  A monitor $\A$ is a \emph{real-time monitor} of statistic $\mu$, if $\semantics{\A} = \mu$.
\end{definition}

However, if the aim of the monitor is to serve as an estimator of the parameter $v$, then it may not be strictly required to output the exact value of $\mu$ at every step, as long as its output almost surely converges to $v$.
A monitor that almost surely converges to $v$ is qualified as \emph{limit}.
Such a monitor is future-oriented, and is concerned with the asymptotic value of the statistic $\mu$ as time tends to infinity, not necessarily computing its precise value over each prefix of the computation.
\begin{definition}[Limit Monitoring]\label{def:limit-monitor}
  A monitor $\A$ is a \emph{limit monitor} of statistic $\mu : \Sigma^* \to \Lambda$ on process $\P$, when $\semantics{\A}(\P) = v$ if and only if $\mu(\P) = v$ for all $v \in \Lambda$.
\end{definition}

In other words, if the statistic converges then the limit monitor converges to the same value, and if the statistic does not converge then neither does the monitor.
To the best of our knowledge, the notion of limit monitoring was not previously considered.
By definition, every real-time monitor is trivially also a limit monitor for the corresponding statistic.
However, in this paper we show that dedicated limit monitors can be much more efficient.
\begin{restatable}{proposition}{propRealTimeImpliesLimit}\label{prop:real-time-implies-limit}
Every real-time monitor of some statistic $\mu$ is also a limit monitor of $\mu$, on arbitrary generating processes.
\end{restatable}
This is in clear contrast to a common trend in runtime verification, where past-oriented monitoring (inherently deterministic) often turns out to be computationally easier than future-oriented monitoring (requiring nondeterministic simulation).


\section{Precise Real-Time Monitoring}\label{sec:precise}
In this section we study the real-time monitoring of statistics by counter monitors.
Real-time monitors can be seen as monitoring the past in a precise manner.
We show that for some common statistics such as the \emph{mode} and \emph{median} statistics this problem is inherently resource-intensive.
More precisely, we identify a class of statistical quantities that require at least as many counters as there are events in the input alphabet.

To illustrate the difficulty of monitoring certain statistics in real time, recall the \emph{mode} as defined in \autoref{ex:mode}.
A straightforward real-time monitor for the mode counts the number of occurrences of each letter $\sigma$ in a separate counter $x_\sigma$.
Then $\sigma$ is the mode if and only if $x_\sigma > x_\rho$ for all $\rho \in \Sigma \setminus \set{\sigma}$.
Hence $|\Sigma|$ counters suffice to monitor the mode.
But can we do better?
Intuitively it seems necessary to keep track of the exact number of occurrences for each individual letter.
Indeed, we show in this section that for real-time monitors this number is tight: any real-time counter monitor of the mode must use at least $|\Sigma|$ counters.
In many applications where the alphabet $\Sigma$ is large this may be beyond the amount of resources available for a monitor.
While \autoref{prop:real-time-implies-limit} implies that the mode can also be limit monitored using $|\Sigma|$ counters, we show in the next section that limit monitoring can be much more resource-sparing.

\looseness=-1
To capture the hardness of real-time monitoring for a whole class of statistics, we start by defining an equivalence relation over words relative to a statistic.
Two words are $\mu$-equivalent if it is impossible for $\mu$ to distinguish them, even with an arbitrary suffix appended to both words.
\begin{definition}[$\mu$-Equivalence]
Let $\mu$ be a statistic over $\Sigma$.
Two words $w_1, w_2 \in \Sigma^*$ are $\mu$-\emph{equivalent}, denoted $w_1 \equiv_{\mu} w_2$, if $\mu(w_1 u) =  \mu(w_2 u)$ for all words $u \in \Sigma^{*}$.
\end{definition}

Now we define the notion of a $\Sigma$-counting statistic, which states that two equivalent words must have exactly the same number of occurrences per letter, modulo a constant shift across all letters.
Intuitively a $\Sigma$-counting statistic induces many equivalence classes, too many to be possibly tracked by a counter monitor with less than $|\Sigma|$ counters.

\begin{definition}[$\Sigma$-Counting]\label{def:sigma-counting}
A statistic $\mu$ is \emph{$\Sigma$-counting} if $w \equiv_\mu w'$ implies that there exists $n \in \integers$ such that $|w|_\sigma = |w'|_\sigma + n$ for all $\sigma \in \Sigma$.
\end{definition}

\begin{restatable}{proposition}{propModeSigmaCounting}\label{prop:mode-sigma-counting}
For any $\Sigma$ such that $|\Sigma|>1$ both the mode and the median statistics are $\Sigma$-counting.
\end{restatable}

To illustrate the definition of $\Sigma$-counting, consider the $\mode$-equivalent words $aabc$ and $a$ over the alphabet $\Sigma = \set{a,b,c}$.
The distance for all letter counts is one.
Over the alphabet with an additional letter $d$ the two words are not $\mode$-equivalent (for example, consider the extensions $aabcd$ and $ad$), since the distance for the count of $d$ is zero.

We prove that $\Sigma$-counting statistics are expensive to monitor by showing that for large $n$ the number of $\mu$-inequivalent words of length at most $n$ is strictly greater than the number of possible configurations reachable by a counter monitor with less than $|\Sigma|-1$ counters over words of length at most $n$.

\begin{restatable}{theorem}{thmSigmaCountingHard}\label{thm:sigma-counting-hard}
Real-time counter monitors of a $\Sigma$-counting statistic require $\Omega(|\Sigma|)$ counters.
\end{restatable}

As a corollary of \autoref{prop:mode-sigma-counting} and \autoref{thm:sigma-counting-hard}, we have that precisely monitoring the mode and the median in real time requires roughly as many counters as the size of the alphabet, which is prohibitive in many practical applications.


\section{Efficient Limit Monitoring}\label{sec:efficient}
In this section we develop a new algorithmic framework for efficient limit monitoring of frequency-based statistics.
We first present a general monitoring scheme and then instantiate it to derive efficient monitoring algorithms for both mode (\autoref{subsec:mode}) and median (\autoref{subsec:median}).
In \autoref{sec:general} we present a monitoring algorithm for a general class of frequency properties.
While corresponding real-time monitors require a number of counters proportional to the size of the input alphabet, our limit monitors only use a constant number of counters (e.g., four for the mode), independent of the alphabet size.
The algorithmic ideas in our monitoring scheme are simple and intuitive, which makes our algorithms easy to understand, implement, and deploy.
However, the correctness proofs are surprisingly hard and required us to develop a new ergodic theory for Markov chains that takes limits over arbitrary subwords (\autoref{subsec:ergodic}).

Our high-level monitoring strategy comprises the following points:
\begin{enumerate}
\item Split the input sequence into subwords of increasing length.
\item In every subword, acquire partial information about the statistic.
\item Assemble global information about the statistic across different subwords.

\end{enumerate}

The idea behind splitting the input sequence into subwords is that when the monitored property involves frequencies of many events, then different events can be counted separately over different subwords, which enables us to reuse registers.
Because of the probabilistic nature of the generator we can still ensure that, in the long run, the monitor value converges to the limit of the statistic.
As we will see, there is great flexibility in how exactly the sequence is partitioned.
In principle, the subwords can overlap or leave gaps arbitrarily, as long as the length of the considered subwords grows ``fast enough''.

\subsection{An Ergodic Theorem over Infixes}\label{subsec:ergodic}

Consider the following Markov chain on the left-hand side, and a random \oword\ generated by this Markov chain in the table on the right-hand side.
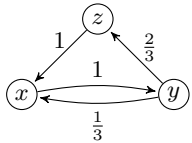
\begin{center}
\small
\begin{minipage}{.22\linewidth}\centering
\begin{tikzpicture}[->,>=stealth',shorten >=1pt,auto]
\node (a) at (0,0) [circle,draw,inner sep=0pt,minimum size=4mm] {$x$};
\node (b) at (2,0) [circle,draw,inner sep=0pt,minimum size=4mm] {$y$};
\node (c) at (1,1) [circle,draw,inner sep=0pt,minimum size=4mm] {$z$};
\path (a) edge [bend left=10] node {1}                (b)
      (b) edge [bend left=10] node {$\frac 1 3$}      (a)
      (b) edge node [above, near start] {$\frac 2 3$} (c)
      (c) edge node [above] {1}                       (a);
\end{tikzpicture}
\end{minipage}%
\begin{minipage}{.78\linewidth}\centering
\setlength\tabcolsep{2pt}
\begin{tabular}{|l||c|cc|ccc|cccc|ccccc|cl|}
\hline
\oword & $x$ & $y$ & $z$ & $x$ & $y$ & $z$ & $x$ & $y$ & $x$ & $y$ & $z$ & $x$ & $y$ & $z$ & $x$ & $y$ & $\dots$ \\
\hline
Prefixes & 0 & .5 & .33 & .25 & .4 & .33 & .29 & .38 & .33 & .4 & .36 & .33 & .38 & .36 & .33 & .38 & $\asto \frac 3 8$ \\
\hline
Infixes & 0 & \multicolumn{2}{r|}{.5} & \multicolumn{3}{r|}{.33} & \multicolumn{4}{r|}{.5} & \multicolumn{5}{r|}{.2} & & $\asto \frac 3 8$ \\
\hline
\end{tabular}
\end{minipage}
\end{center}
The second row of the table shows the frequency of state $y$ in prefixes of increasing length.
For example, after $xyzx$ we have frequency $\frac 1 4$.
The classic ergodic theorem (\autoref{thm:ergodic-classic}) tells us that this frequency almost surely converges to $f_y = \frac 3 8$, the inverse of the expected return time to $y$.
However, this theorem does not apply to take a limit over arbitrary subwords, for example, the infixes of increasing length (indicated by vertical lines) in the third row of the table.
We prove a result that shows that also in this much more general case the limit frequency of $y$ is $\frac 3 8$.

The strong law of large numbers states that the empirical average of \iid\ random variables converges to their expected value, i.e., $(X_1 + \dots + X_n) / n \asto \EE(X_1)$ as $n \to \infty$.
The fact that random variables are ``reused'' from the $n$'th to the $(n+1)$'st sample does matter in this statement. Otherwise the mere existence of a mean value is not sufficient to guarantee convergence. However, when the variance (or higher-order moment) is bounded, then this ``reuse'' is no longer required. We now prove such a variant of the law of large numbers.\footnote{Such a setting is sometimes called array of rowwise independent random variables in the literature, see~\cite{HuMT89} in particular.}
\begin{restatable}{theorem}{thmGslln}\label{thm:gslln}
  Let $\set{X_{n,i} : n,i \ge 1}$ be a family of identically distributed random variables with
  $\EE(X_{1,1}) = \mu$ and $\EE(X_{1,1}^4) < \infty$,
  such that $\set{X_{n,i} : i \ge 1}$ are mutually independent for every $n \ge 1$.
  Let $(s_n)_{n \ge 1}$ be a sequence of indices with $s_n \ge an$ for every $n \ge 1$ and fixed $a > 0$.
  Set $S_n = \sum_{i=1}^{s_n} X_{n,i}$.
  Then
  $S_n / s_n \asto \mu$ as $n \to \infty$.
\end{restatable}
In our proof the combination of the fourth-moment bound and the linear increase of $s_n$ leads to a converging geometric series.
We believe that these assumptions could be slightly relaxed to a second-moment bound or to sublinearly increasing sequences.
\autoref{thm:gslln} already gives a basis to reason about infix-convergence for \iid\ processes.
We now use it to derive a corresponding result for Markov chains.

Let $\M$ be a Markov chain and $(X_i)_{i \ge 1}$ be $\mathit{Markov}(\M)$.
Given an \emph{offset function} $\offset: \naturals \to \naturals$, we refer to $X_{\offset(n)+1} X_{\offset(n)+2} \cdots$ as the \emph{$n$'th suffix} of $X$.
We denote by $\mcV(k) = \sum_{i=1}^{k} 1_{\set{X_{\offset(n)+i} = q}}$ the \emph{number of visits} to state $q$ within $k$ steps in the $n$'th suffix.
We generalize the classic ergodic theorem for Markov chains (\autoref{thm:ergodic-classic}) to take the limit over arbitrary subwords.
\begin{restatable}
{theorem}{thmErgodic}\label{thm:ergodic}
  Let $\M$ be a finite connected Markov chain and $\offset$ an offset function.
  If $(X_i)_{i \ge 1}$ is $\mathit{Markov}(\M)$ then
  ${\mcV(n)} / n \asto \freq_q$ as $n \to \infty$
  for every state $q$.
\end{restatable}
Our proof applies \autoref{thm:gslln} to the \iid\ excursion times between visiting state $q$ within the $n$'th suffix.
This requires bounding the moments of excursion times and showing that the time until visiting $q$ for the first time in every subword becomes almost surely negligible for increasing size subwords.
As a corollary of \autoref{thm:ergodic} we get the following characterization for the long-run frequencies of letters over infixes.
\begin{corollary}\label{cor:ergodic}
  Let $\M$ be a finite connected Markov chain and $\offset$ an offset function.
  If $w$ is $\mathit{Markov}(\M)$ then
  ${|w_{\offset(n)+1..\offset(n)+n}|_\sigma} / n \asto \freq_\sigma$ as $n \to \infty$
  for every letter $\sigma$.
\end{corollary}

\subsection{Monitoring the Mode}\label{subsec:mode}

As we saw in \autoref{sec:precise}, precisely monitoring the mode in real time requires at least $|\Sigma|$ counters.
By contrast, we now show that the mode can be limit monitored using only four counter registers.
For convenience we also use two registers to store event letters; since we assume $\Sigma$ to be finite they can be emulated in the finite state component of the monitor.

The core idea of our monitoring algorithm is to split $w$ into chunks, and for each chunk only count the number of occurrences of two letters $x$ and $y$.
Letter $x$ is considered the current candidate for the mode and $y$ is a randomly selected contender.
If $x$ does not occur more frequently than $y$ in the current chunk, $y$ becomes the mode candidate for the next chunk.
The success of the monitor relies on two points: (i)~it must be repeatably possible for the true mode to end up in $x$, and (ii)~it must be likely for the true mode to eventually remain in $x$.
The first point is achieved by taking $y$ randomly, and the second point is achieved by gradually increasing the chunk size.
It is sufficient to increase the chunk size by one and decompose $w$ as follows:
\begin{align*}
  \underbracket{\, \sigma_1 \,}\,
  \underbracket{\, \sigma_2 \sigma_3 \,}\,
  \underbracket{\, \sigma_4 \sigma_5 \sigma_6 \,}\,
  \underbracket{\, \sigma_7 \sigma_8 \sigma_9 \sigma_{10} \,}\,
  \underbracket{\, \sigma_{11} \cdots\,}
\end{align*}
Formally, the decomposition of $w$ into chunks is given by an offset function $\offset : \naturals \to \naturals$ with $\offset(n) = \frac{n(n-1)}{2}$, such that the $n$'th chunk starts at $s(n)+1$ and ends at $s(n)+n$.
For convenience, we introduce a double indexing of $w$ by $n \ge 1$ and $1 \le i \le n$, such that $w_{n,i} = w_{s(n)+i}$ is the $i$'th letter in the $n$'th chunk.

\begin{figure}[tb]
  \begin{minipage}{.48\linewidth}
    \begin{algorithm}[H]\small
      \caption{Mode monitor}
      \label{alg:mode}

      \SetKwFunction{Init}{Init}
      \SetKwFunction{Next}{Next}
      
      \Function{\Init{$\sigma$}}{
        $x,y \assign \sigma,\sigma$\;\label{alg:mode:init}
        $c_x,c_y \assign 0,0$\;
        $n,i \assign 2,1$\;
        \KwRet $x$\;
      }
      \Function{\Next{$\sigma$}}{
        \If{$i = 1$}{
          \lIf{$c_x \le c_y$}{
            $x \assign y$\label{alg:mode:x}
          }
          $y \assign \sigma$\;\label{alg:mode:y}
          $c_x,c_y \assign 0,0$\;
        }
        \;
        \lIf{$x = \sigma$}{$c_x \assign c_x + 1$}\label{alg:mode:cnt1}
        \lIf{$y = \sigma$}{$c_y \assign c_y + 1$}\label{alg:mode:cnt2}
        \;
        \lIf{$i = n$}{$n,i \assign n+1,1$}\hspace{12.5mm}\lElse{$i \assign i+1$}
        \KwRet $x$\;
      }
    \end{algorithm}
  \end{minipage}\hfill
  \begin{minipage}{.48\linewidth}
    \begin{algorithm}[H]\small
      \caption{Median monitor}
      \label{alg:median}

      \SetKwFunction{Init}{Init}
      \SetKwFunction{Next}{Next}
      
      \Function{\Init{$\sigma$}}{
        $x \assign \sigma$\;
        $c_1,c_2,c_3,c_4 \assign 0,0,0,0$\;
        $n,i \assign 2,1$\;
        \KwRet $x$\;
      }
      \Function{\Next{$\sigma$}}{
        \If{$i = 1$}{
          \lIf{$c_1 \ge c_2$}{$x \assign \mathit{pre}_\slt(x)$}\label{alg:median:pre}
          \lIf{$c_3 \ge c_4$}{$x \assign \mathit{succ}_\slt(x)$}\label{alg:median:succ}
          $c_1,c_2,c_3,c_4 \assign 0,0,0,0$\;
        }
        \lIf{$\sigma < x$}{$c_1 \assign c_1 + 1$}\label{alg:median:cnt-start}
        \lIf{$\sigma \ge x$}{$c_2 \assign c_2 + 1$}
        \lIf{$\sigma > x$}{$c_3 \assign c_3 + 1$}
        \lIf{$\sigma \le x$}{$c_4 \assign c_4 + 1$}\label{alg:median:cnt-end}
        \lIf{$i = n$}{$n,i \assign n+1,1$}\hspace{12.5mm}\lElse{$i \assign i+1$}
        \KwRet $x$\;
      }
    \end{algorithm}
  \end{minipage}
\end{figure}

A formal description of our mode monitor is given in \autoref{alg:mode}.
The counters $n$ and $i$ keep track of the decomposition of $w$.
For the very first letter $\sigma$, \texttt{Init} initializes both registers $x$ and $y$ to $\sigma$ (\autoref{alg:mode:init}).
Then, for every subsequent letter, \texttt{Next} counts an occurrence of $x$ and $y$ using counters $c_x$ and $c_y$, respectively (\autoref{alg:mode:cnt1}-\ref{alg:mode:cnt2}).
At the beginning of every chunk, $x$ is replaced by $y$ if it did not occur more frequently in the previous chunk (\autoref{alg:mode:x}), and $y$ is set to the first letter of the chunk (\autoref{alg:mode:y}).
At every step, $x$ is the current estimate of the mode.

\begin{example}\label{ex:mode-iid}
For alphabet $\Sigma = \set{a,b,c}$ and probability distribution $p$ with $p(a)=0.5$, $p(b)=0.3$, and $p(c)=0.2$,
the following table shows a word $w$ where every letter was independently sampled from $p$, and the corresponding mode at every position in $w$.
\begin{center}
  \begin{tabular}{|c|c|}
    \hline
    $w$ & c b b a b a c a a b c a c a a a $\cdots$ \\
    $\mode$    & c -\, b b b b b - a -\, -\, a a a a a $\cdots$ \\
    \hline
  \end{tabular}
\end{center}
In this example, $\mode$ first switches between the different letters and undefined, but then eventually seems to settle on $a$.
We show that this is not an accident, but happens precisely because $a$ is the unique letter that $p$ assigns the highest probability.

Now the following table shows the execution of \autoref{alg:mode} on the same random word.
\begin{center}
  \newcommand{\al}[1]{\multicolumn{1}{l|}{#1}}
  \begin{tabular}{|c||c|c|c|c|c|c|}
    \hline
    $n$      & 1 & \al 2 & \al 3 & \al 4   & \al 5     & 6 $\cdots$ \\
    $i$      & 1 & 1 2   & 1 2 3 & 1 2 3 4 & 1 2 3 4 5 & 1 $\cdots$ \\\hline
    $\sigma$ & c & b b   & a b a & c a a b & c a c a a & a $\cdots$ \\\hline
    $x$      & c & \al c & \al b & \al a   & \al a     & a $\cdots$ \\
    $y$      & c & \al b & \al a & \al c   & \al c     & a $\cdots$ \\
    $c_x$    & 1 & 0 0   & 0 1 1 & 0 1 2 2 & 0 1 1 2 3 & 1 $\cdots$ \\
    $c_y$    & 1 & 1 2   & 1 1 2 & 1 1 1 1 & 1 1 2 2 2 & 1 $\cdots$ \\
    \hline
  \end{tabular}
\end{center}
Initially $c$ is considered the mode and compared to $b$ in the second chunk, where $b$ occurs more frequently.
Thus $b$ is considered the mode and compared to $a$ in the third chunk, where $a$ occurs more frequently.
In the fourth and fifth chunk $a$ is compared to $c$, where $a$ occurs more frequently in both chunks.
Again, the algorithm seems to settle on $a$, the true mode.
\end{example}

To prove the correctness of our algorithm according to \autoref{def:limit-monitor} requires us to first characterize when a Markov chain has a mode, i.e., under which conditions the mode statistic almost surely converges.
For this it is illustrative to instantiate \autoref{def:convergence} for the mode, which states that $a$ is the mode of an \oword\ $w$ if there exists a length $n$, such that for every length $n' \ge n$, $|w_{..n'}|_a > |w_{..n'}|_b$ for every $b \neq a$.
In a Markov chain the ergodic theorem characterizes the long-run frequencies of states, and thus the long-run frequencies of letters (see \autoref{cor:ergodic-classic}).
Hence a Markov chain has a mode if and only if its random \oword\ almost surely has a unique letter that occurs most frequently.
\begin{restatable}{theorem}{thmModeMarkov}\label{thm:mode-markov}
  Over Markov chains, the mode statistic converges to $a$ if and only if $\freq_a > \freq_b$ for all $b \neq a$.
\end{restatable}
\begin{proof}
  Let $\M$ be a Markov chain and $w$ be $\Markov(\M)$.
  According to \autoref{cor:ergodic-classic}, 
  ${|w_{..n}|_\sigma}/{n} \asto \freq_\sigma$ as $n \to \infty$
  for every $\sigma \in \Sigma$
  
  Now assuming $\freq_a > \freq_b$ for all $b \neq a$, we have for sufficiently large $n$ that $|w_{..n}|_a > |w_{..n}|_b$ for all $b \neq a$, and thus $a$ is the mode of $w$ almost surely.

  Conversely, if there are two distinct letters $a,a'$ with equal maximal frequencies $\freq_a, \freq_{a'}$, then almost surely the mode switches infinitely often between $a$ and $a'$, thus neither $a$ nor $a'$ is the mode of $w$, and thus $w$ does not have a mode.
\end{proof}

Now we can prove that \autoref{alg:mode} is a limit monitor for the mode.
The core of the argument is that the probability of the true mode eventually staying in register $x$ is lower-bounded by the probability of $a$ eventually being the most frequent letter in \emph{every} subword and $a$ being eventually selected into $y$, which happens almost surely.

\begin{restatable}{theorem}{thmModealgoMarkov}\label{thm:modealgo-markov}
  \autoref{alg:mode} limit-monitors the mode over Markov chains.
\end{restatable}
\begin{proof}
  Let $w$ be $\Markov(\M)$ and let $a$ be the mode of $w$ (the other case where $w$ does not have a mode is obvious).
  Let $\infixCnt_n$ be the function that maps every letter to the number of its occurrences in the $n$'th subword, i.e., $\infixCnt_n(\sigma) = |w_{\offset(n)+1..\offset(n)+n}|_\sigma$.
  To capture \autoref{alg:mode} mathematically, we define the random variables
  \begin{align*}
    Y_n &= w_{n,1}; &
    X_1 &= w_{1,1}; &
    X_{n+1} &=
          \twopartdef
            {X_{n},}{\infixCnt_n(X_{n})  >  \infixCnt_n(Y_{n});}
            {Y_{n},}{\infixCnt_n(X_{n}) \le \infixCnt_n(Y_{n}).}
  \end{align*}
  That is, $X_n$ and $Y_n$ are the values of $x$ and $y$ throughout the $n$'th subword.
  We need to show that almost surely, eventually $X_n = a$ forever, i.e., $\PP(\eventually\always X_n = a) = 1$.\footnote{In the interest of readability we use temporal (modal) logic notation $\eventually$ and $\always$ meaning \emph{eventually} and \emph{forever}, respectively.}
  
  It is more likely that $a$ eventually stays in $x$ forever as that $a$ eventually is the most frequent letter in \emph{every} subword and that $a$ is also eventually sampled into $y$:
  \begin{align*}
    &\phantom{{}\ge{}} \PP(\eventually\always X_n = a) \\
    &\ge \PP(\eventually(\always \forall b \neq a : \infixCnt_n(b) < \infixCnt_n(a)) \land (\eventually Y_n = a)) \\
    &\ge \PP((\always_{\ge n_0} \forall b \neq a : \infixCnt_n(b) < \infixCnt_n(a)) \land (\eventually_{\ge n_0} Y_n = a))
  \end{align*}
  The last lower bound holds for any fixed $n_0$ and we show that it converges to 1 as $n_0 \to \infty$.
  \begin{align*}
    &\phantom{{}\ge{}} \PP((\always_{\ge n_0} \forall b \neq a : \infixCnt_n(b) < \infixCnt_n(a)) \land (\eventually_{\ge n_0} Y_n = a)) \\
    &\ge \PP(\always_{\ge n_0} \forall b \neq a : \infixCnt_n(b) < \infixCnt_n(a)) \cdot \PP(\eventually_{\ge n_0} Y_n = a)\\
    &= \PP(\always_{\ge n_0} \forall b \neq a : \infixCnt_n(b) < \infixCnt_n(a))
  \end{align*}
  Since $\infixCnt_n(\sigma) / n \asto \freq_\sigma$ (by \autoref{cor:ergodic}) and $a$ is the unique letter with highest frequency $\freq_a$ (by \autoref{thm:mode-markov}), we have $\PP(\always_{\ge n_0} \forall b \neq a : \infixCnt_n(b) < \infixCnt_n(a)) = 1$ for sufficiently large $n_0$. Thus, $\PP(\eventually\always X_n = a) = 1$.
\end{proof}

Note that our policy of always selecting the mode contender $y$ from the input is an optimization, since we expect to see the mode often in the input.
Our proof requires that the true mode is selected into $y$ infinitely often, which is the case because we update $y$ at irregular positions.
Two other policies to update $y$ would be (i)~to always uniformly sample from $\Sigma$, or (ii)~to cycle deterministically through all elements of $\Sigma$.

\subsection{Monitoring the Median}\label{subsec:median}

Recall from \autoref{ex:median} that $a$ is the \emph{median} of a word $w$ over a $\slt$-ordered alphabet $\Sigma$ when
\begin{equation}\label{eq:le}
\sum_{\sigma \sgt a} |w|_\sigma < \sum_{\sigma \sle a} |w|_\sigma
\end{equation}
on the one hand, and
\begin{equation}\label{eq:ge}
\sum_{\sigma \slt a} |w|_\sigma < \sum_{\sigma \sge a} |w|_{\sigma}
\end{equation}
on the other hand.
These equations readily lead to our median limit-monitoring algorithm shown in \autoref{alg:median}, which we display next to our mode monitor to highlight their common structure.
The idea of the algorithm is to maintain a median candidate $x$ and then use four counters $c_1,c_2,c_3,c_4$ to compute the sums in inequality~\eqref{eq:le} and~\eqref{eq:ge}, for $a=x$, in every subword (\autoref{alg:median:cnt-start}-\ref{alg:median:cnt-end}).
Whenever any of the two inequalities is not satisfied at the end of a subword, a new median candidate is selected into $x$ for the next subword.
In particular, if inequality~\eqref{eq:le} is violated then the next lower value in the ordering $\slt$ is selected (\autoref{alg:median:pre}), and if inequality~\eqref{eq:ge} is violated then the next higher value is selected (\autoref{alg:median:succ}).
Notice that we could eliminate the counters $c_3,c_4$, by alternating the computation of inequality~\eqref{eq:le} and~\eqref{eq:ge} over different subwords, and thus reusing $c_1,c_2$ to compute inequality~\eqref{eq:ge}.

\begin{restatable}{theorem}{thmMedianMarkov}\label{thm:medianalgo-markov}
  \autoref{alg:median} limit-monitors the median over Markov chains.
\end{restatable}


\section{Monitoring General Frequency Properties}\label{sec:general}
In the previous section we presented high-level principles for efficient limit monitoring and designed specialized monitoring algorithms for the mode and median statistic, which are both derived from event frequencies.
We postulate that our algorithmic ideas are straightforward to adapt to obtain monitors for many other frequency-based statistics.
However, we did not yet precisely define what we mean by \emph{frequency property}, nor demonstrated how efficiently these can be limit monitored in general.
In this section we provide a first step in this direction by defining a simple language to specify frequency-based Boolean statistics, and showing that all statistics definable in this language can be limit monitored over Markov chains with four counters only.

From the defining equations of the mode and median we observe that a characteristic construction is the formation of linear inequalities over the frequencies (or equivalently, occurrence counts) of specific events.
The key part of the argument for the correctness of our monitoring algorithms is that since event frequencies almost surely converge, both over prefixes and infixes, also these inequalities almost surely ``stabilize''.
We use the same construction at the core of a language to define general frequency-based statistics.
For simplicity we focus on statistics that output a Boolean value.

\begin{definition}
A \emph{frequency formula} over alphabet $\Sigma$ is a Boolean combination of atomic formulas of the form
\begin{align}
  \sum_{\sigma \in \Sigma} \alpha_\sigma \cdot \freq_\sigma > \alpha \label{eq:atom}
\end{align}
where all $\alpha$'s are integer coefficients.
\end{definition}
A frequency formula $\phi$ is built from linear inequalities over frequencies of events.
The evaluation of a frequency formula is as expected (we write $w \models \phi$ if $\phi$ evaluates to true over $w$).
Hence we see $\phi$ as defining the Boolean statistic $\semantics{\phi} : \Sigma^* \to \booleans$, where
\begin{align*}
  \semantics{\phi}(w) = \twopartdef{1,}{w \models \phi;}{0,}{w \not\models \phi.}
\end{align*}

\begin{example}
The existence of a mode is expressed as the frequency formula
\begin{align*}
  \bigvee_{a \in \Sigma} \bigwedge_{\substack{\sigma\in\Sigma \\ \sigma \neq a}} f_a > f_\sigma \,.
\end{align*}
\end{example}

\begin{example}
Consider a web server that favors certain client requests over others.
Such a malfunction could be observed by detecting that certain events are disproportionately more frequent than others.
The following frequency formula specifies that no event can occur 100-times more frequent than any other event:
\begin{align*}
  \bigwedge_{\substack{a,b \in \Sigma \\ a \neq b}} f_a < 100 \cdot f_b \,.
\end{align*}
\end{example}

A frequency formula $\phi$ can be limit monitored by simply evaluating $\phi$ repeatedly over longer and longer subwords.
However, the key to save resources is to evaluate different atomic subformulas of $\phi$ over different subwords, and thus only evaluating one subformula at a time.
\begin{theorem}
Over Markov chains, every frequency formula can be limit monitored using 4 counters.
\end{theorem}
\begin{proof}[Proof]
Let $\phi$ be a frequency formula with $k$ atomic subformulas $\phi_1,\dots,\phi_k$ of the form \eqref{eq:atom}.
The monitor partitions the input word $w$ into infixes $w_{n,i}$ with $|w_{n,i}| = n$, for $n \ge 1$ and $1 \le i \le k$, as follows:
\begin{align*}
  \dots
  \underbracket{
    \underbracket{w_{n,1}}_{\phi_1}
    \underbracket{w_{n,2}}_{\phi_2}
    \dots
    \underbracket{w_{n,k}}_{\phi_k}
  }_{\phi}
  \dots
\end{align*}
Keeping track of the increasing infix length $n$ and the current position within an infix requires two counters.
Then over every infix $w_{n,i}$ the monitor uses two counters to compute $\phi_i$, one for positive and one for negative increments.
At the end of $w_{n,i}$ we have a truth value for $\phi_i$ that is used to partially evaluate $\phi$.
This evaluation is implemented in the final-state component of the monitor, and the two counters are reused across all infixes.
Then after every $k$'th infix we have a new ``estimate'' of $\phi$ that in the long run converges the same way as $\semantics{\phi}$.
Hence the resulting automaton is a limit monitor of $\phi$: by~\autoref{cor:ergodic}, the frequency of each event over infixes of increasing length tends to its respective asymptotic frequency, so that strict inequalities holding over empirical frequencies almost surely hold over infixes of increasing length.
\end{proof}


\section{Conclusion}\label{sec:conclusion}
In this paper we have studied the monitoring of frequency properties of event sequences.
We observed that real-time monitoring can be surprisingly hard (i.e., resource-intensive) for such properties, and introduced the alternative notion of limit monitoring.
In this limit-monitoring setting we showed that a simple algorithmic idea leads to resource-efficient monitoring algorithms for frequency properties.
To prove the correctness of our algorithms we generalized the ergodic theory of Markov chains.

The results in this paper are a first indicator of the relevance and potential of limit monitoring.
We hope that future research broadens the understanding of this problem and we close with a number of interesting directions.

First, we are interested in a tighter characterization of properties that can be efficiently limit monitored.
Let us remark that the results in this paper immediately generalize from counting individual events to counting the occurrences of regular event patterns.
This is the case because regular expression matching can be performed in real time by the finite state component of a counter monitor.
We extended our frequency formulas with free variables to support non-Boolean statistics, and quantification to reason about unknown alphabet symbols.
However, the shape and efficiency of a generic monitoring algorithm is not yet clear.
For examples, we saw that there are different policies to partition the input sequence and different policies to obtain candidate values for the monitor output.
Certain forms of existential quantification can be translated to random sampling, but this does not seem to hold in general since not all events in the alphabet may occur in the execution under consideration.
Going even further, it would be interesting to consider limit monitoring of properties with temporal aspects (such as \emph{always} and \emph{eventually} modalities).

Second, it is well known (see e.g.~\cite{Booth68}) that the asymptotic frequencies of $k$-long subwords fully characterize a $k$-state connected Markov chain.
Hence the transition probabilities of a Markov chain (of known structure) can be inferred from the conditional probabilities of events.
Thus, assuming the structure of a Markov chain is known, frequency queries and the algorithmic ideas in this paper can be used to learn its transition probabilities to an arbitrary precision.
It would be interesting to study more broadly ``how much'' of a system can be learned from frequency properties (and similar observations).

Third, throughout this paper we used the term efficient to mean resource-efficient in the amount of memory used by a monitor.
However, there is the orthogonal question of time-efficiency.
For a limit monitor this means how quickly a monitor converges in relation to the monitored statistic.
We hope that future research can provide numerical guarantees or estimates for convergence rates.
For the simple setting of an \iid\ word over a two-letter alphabet, we proved that the mode statistic converges exponentially fast.
More precisely, if $w$ is a random \oword\ where every letter is \iid\ according to a probability distribution $p$ over $\set{a,b}$ with $p(a) > p(b)$, then $\PP(\mode(w_{..n}) = a) \ge 1 - (4p(a)p(b))^{\lfloor \frac n 2 \rfloor}$.
Since this depends on the exact probabilities, the analytical expressions of the confidence value seem to become intractable for three letters or more.
In probability theory, there exist several different notions of convergence of random variables.
The results in this paper use the notion of almost-sure convergence of a statistic $\mu$ (\autoref{def:convergence}), that is, $\PP_{w\sim \P}(\lim_{n \to \infty} \mu(w_{n..}) = v) = 1$.
It would be interesting to study also other notions, for example convergence in probability, that is, $\lim_{n \to \infty} \PP_{w\sim \P}(\mu(w_{n..}) = v) = 1$.

Fourth, the correctness results we derived for our monitoring algorithms hold for systems modeled as connected Markov chains.
However, we believe that the algorithmic ideas of this paper are more widely applicable.
Thus it would be interesting to study limit monitoring for other types of systems, for example, Markov decision processes which are challenging for our monitoring scheme because nondeterminism allows certain events to \emph{always} occur deliberately when the monitor is not watching for them.
In the security context a monitored system is usually assumed to be adversarial, not probabilistic.
It could be interesting to turn our deterministic monitors of probabilistic systems into probabilistic monitors for nondeterministic systems.


\bibliographystyle{plainurl}
\bibliography{references}

\appendix

\section{Proofs of \autoref{sec:monitoring}}

\propRealTimeImpliesLimit*

\begin{proof}
Let $\A$ be a real-time monitor of $\mu$. We have $\semantics\A = \mu$, hence $\semantics\A$ converges to some value $v$ over a generating process $\P$ if and only if $\mu$ converges to $v$ over $\P$.
\end{proof}

\section{Proofs of \autoref{sec:precise}}

\propModeSigmaCounting*

\begin{proof}
We prove the statement for the case of the mode.
Let $w$ and $w'$ be words such that $w \equiv_\mode w'$. 
Assume, towards a contradiction, that for all $n \in \integers$ there exists $\sigma \in \Sigma$ such that $|w|_\sigma \neq |w'|_\sigma + n$.
By assumption $w$ and $w'$ have the same mode $\rho$, otherwise they are trivially not equivalent according to $\equiv_\mode$.
Let $k = |w|_\rho - |w'|_\rho$.
There exists $\sigma \in \Sigma$ such that $|w|_\sigma \neq |w'|_\sigma + k$.
We have in particular $\sigma \neq \rho$.
Let $l = |w|_\rho - |w|_\sigma$ and $l' = |w'|_\rho - |w'|_\sigma$.
But then $\mode(w\sigma^{l}) \neq \mode(w'\sigma^{l})$ or $\mode(w\sigma^{l'}) \neq \mode(w'\sigma^{l'})$, which contradicts $w \equiv_\mode w'$.
\end{proof}

\thmSigmaCountingHard*

\begin{proof}
Let $\Sigma = \set{\sigma_1,\dots,\sigma_k}$ be a finite alphabet of size $k \ge 2$, $\mu$ be a $\Sigma$-counting statistic, and $\A$ be a $k'$-counter monitor with $m$ states.
We show that for $k' < k-1$ and large enough $n$ the number of $\mu$-inequivalent words of length at most $n$ is strictly greater than the number of possible configurations reachable by $\A$ over words of length at most $n$.

Consider the language $L = \set{\sigma_2^{i_2} \dots \sigma_k^{i_k} \mid 0 \le i_2,\dots,i_k \le n}$.
Observe that all words in $L$ are $\mu$-inequivalent and thus there are at least $(n+1)^{k-1}$ equivalence classes of $\equiv_\mu$ over words of length at most $n (k-1)$.
We assume without loss of generality that $\A$ increments counters by at most one at every event~\cite{FischerMR68}.
There are $m(\ell+1)^{k'}$ possible configurations of $\A$ over words of length at most $\ell$.
For $\ell = n (k-1)$ we bound the number of configurations from above by $(mk^{k'})n^{k'}$.
But then we have that $(n+1)^{k-1} > (mk^{k'})n^{k'}$ for $k' < k-1$ and sufficiently large $n$.
Hence for some integer $n$, there are more $\mu$-inequivalent words of length at most $n$ than configurations that $\A$ can possibly reach after reading such words.
It follows that there is no real-time counter monitor for $\mu$ with less than $k-1$ counters.
\end{proof}

\section{Proofs of \autoref{sec:efficient}}

\thmGslln*

\begin{proof}
  Let $\EE(X_{1,1}^4) = M$.
  W.l.o.g.~$\mu = 0$ (consider $Y_{n,i} = X_{n,i} - \mu$).
  We expand $\EE(S_n^4)$ and observe that, by independence,
  $ \EE(X_{n,i} X_{n,j}^3) = \EE(X_{n,i} X_{n,j} X_{n,k}^2) = \EE(X_{n,i} X_{n,j} X_{n,k} X_{n,l}) = 0 $
  for distinct indices $i,j,k,l$.
  Hence,
  \begin{align*}
    \EE(S_n^4)
    &= \EE\left( \sum_{1 \le i \le s_n} X_{n,i}^4 + 6 \sum_{1 \le i < j \le s_n} X_{n,i}^2 X_{n,j}^2 \right) \,.
  \end{align*}
  Now for $i \le j$, by independence and the Cauchy-Schwarz inequality
  \[ \EE(X_{n,i}^2 X_{n,j}^2) = \EE(X_{n,i}^2) \EE(X_{n,j}^2) \le \EE(X_{n,i}^4)^{\frac 1 2} \EE(X_{n,j}^4)^{\frac 1 2} = M \,. \]
  So we get the bound
  \[ \EE(S_n^4) \le s_nM + 3s_n(s_n-1)M \le 3s_n^2M \,. \]
  Thus
  \[ \EE\left( \sum_{n \ge 1} (S_n/s_n)^4 \right) \le 3M \sum_{n \ge 1} 1/s_n^2 \le \tfrac{3M}{a^2} \sum_{n \ge 1} 1/n^2 < \infty \]
  which implies
  \[ \sum_{n \ge 1} (S_n/s_n)^4 < \infty \text{ a.s.} \]
  and hence $S_n/s_n \asto 0$.
\end{proof}

\thmErgodic*

To prove the result, we first introduce some notation and supporting lemmas.

We denote by $\mcT r$ (for $r \ge 0$) the $r$'th time of visiting state $q$ in the $n$'th subword, and by $\mcS r$ (for $r \ge 1$) the length of the $r$'th excursion to state $q$ in the $n$'th subword:
\begin{align*}
  \mcT 0 &= \inf\set{i \ge 1 \mid X_{\offset(n)+i} = q}; \\
  \mcT {r+1} &= \inf\set{i > \mcT {r} \mid X_{\offset(n)+i} = q}; \\
  \mcS {r+1} &= \mcT {r+1} - \mcT r .
\end{align*}
Let $\mcSS(k)$ be the length of the first $k$ excursions to state $q$, and $\mcTS(k)$ additionally includes the time to visit $q$ for the first time:
\begin{align*}
  \mcSS(k) = \sum_{i=1}^k \mcS i; \qquad \mcTS(k) = \mcT 0 + \mcSS(k) \,.
\end{align*}

For a state $q$ we establish the following connection between the time it takes to visit $q$ a certain number of times, and the number of times $q$ is visited within a certain time bound.
\begin{restatable}{lemma}{lemExchange}\label{lem:exchange}
  For $a \ge 0$  and arbitrary $b$, we have
  \begin{align}
    \mcTS(k) \le n + a &\implies \mcV(n) \ge k - a   \,; \label{eq:mcTSV1} \\
    \mcTS(k) \ge n - b &\implies \mcV(n) \le k + 1 + |b| \,. \label{eq:mcTSV2}
  \end{align}
\end{restatable}
\begin{proof}
  $\mcTS(k)$ is the time of visiting $q$ for the $(k+1)$'th time.
  In~\eqref{eq:mcTSV1} this is at most $\floor a$ steps beyond $n$.
  If we walk back $\floor a$ steps to be within $n$, then at worst every step is a $q$ and thus $\mcV(n) \ge k + 1 - \floor a \ge k - a$.
  In~\eqref{eq:mcTSV2}, for $b \ge 0$, $\mcTS(k)$ is at most $\floor b$ steps before $n$.
  If we walk forward $\floor b$ steps to be beyond $n$, we can visit $q$ at most $\floor b$ more times before crossing $n$ and thus $\mcV(n) \le k + 1 + \floor b \le k + 1 + b$.
  The case $b < 0$ is trivial.
\end{proof}

As final ingredients we need that the moments of the recurrence times are finite, and that the ``setup time'' to visit state $q$ for the first time is negligible over increasing length infixes.
\begin{restatable}{lemma}{lemMcMoments}\label{lem:mc-moments}
  The moments of $\mcT 0$ and $\mcS r$ are finite.
\end{restatable}
\begin{proof}
  By Karamata's Tauberian theorem, the explicit formulas for the recurrence moments derived in~\cite{Szewczak08} converge for finite Markov chains.
\end{proof}

\begin{restatable}{lemma}{lemTtoZero}\label{lem:T-to-zero}
  $\mcT 0 / n \asto 0$ as $n \to \infty$.
\end{restatable}
\begin{proof}
  By \autoref{lem:mc-moments} the second moments of $\mcT 0$ are finite, and because there are finitely many states there is a constant $C$ such that $\EE((\mcT 0)^2) \le C$ for all $n \ge 1$.
  Thus
  \[ \EE\left( \sum_{n \ge 1} (\mcT 0 / n)^2 \right) \le C \sum_{n \ge 1} 1/n^2 < \infty \]
  which implies
  \[ \sum_{n \ge 1} (\mcT 0 / n)^2 < \infty \text{ a.s.} \]
  and hence $\mcT 0 / n \asto 0$.
\end{proof}

Now we are ready to prove our generalized ergodic theorem.
\begin{proof}[Proof of \autoref{thm:ergodic}]
  For every $n$, the $\mcS r$'s are \iid\ with expected value $\mcm$.
  Thus, by \autoref{thm:gslln},
  \begin{align*}
    {\mcSS(\ceil{\tfrac n \mcm})}/{\ceil{\tfrac n \mcm}} \asto \mcm \as n \to \infty,
  \end{align*}
  i.e., almost surely
  \begin{align*}
    \forall \epsilon>0 \exists \delta \forall n>\delta : \mcm-\epsilon \le \mcSS(\ceil{\tfrac n \mcm}) / \ceil{\tfrac n \mcm} \le \mcm+\epsilon \,.
  \end{align*}
  Inside the quantifiers we multiply $\ceil{\tfrac n \mcm}$, add $\mcT 0$, and derive
  \begin{align*}
    n - (\tfrac{n\epsilon}{\mcm} + \epsilon - \mcT 0)
    \le \mcTS(\ceil{\tfrac n \mcm}) \le
    n + (\tfrac{n\epsilon}{\mcm} + \mcm + \epsilon + \mcT 0) \,.
  \end{align*}
  By applying \autoref{lem:exchange} and dividing by $n$ we obtain
  \begin{align*}
    \tfrac 1 \mcm - \left( \tfrac \epsilon m + \tfrac{m+\epsilon}{n} + \tfrac{\mcT 0}{n} \right)
    \le \mcV(n) / n \le
    \tfrac 1 \mcm + \left( \tfrac \epsilon m + \tfrac{2+\epsilon}{n} + \tfrac{\mcT 0}{n} \right) \,.
  \end{align*}
  By \autoref{lem:T-to-zero} and suitably chosen $\epsilon$, the terms in parenthesis can be made arbitrarily small for sufficiently large $n$.
  Thus, for any given $\epsilon'>0$, almost surely $\frac 1 \mcm - \epsilon' \le \mcV(n) / n \le \frac 1 \mcm + \epsilon'$ for sufficiently large $n$, proving the theorem.
\end{proof}

\section{Convergence Rate of the Mode}\label{app:convergence-rate}

\begin{proposition}
Let $p$ be a probability distribution over the alphabet $\set{a,b}$ with $p(a) > p(b)$.
Let $w$ be a random \oword\ where every letter is \iid\ according to $p$.
Then $\PP(\mode(w_{..n}) = a) \ge 1 - \rho^{\lfloor \frac n 2 \rfloor}$, for $\rho = 1-(2p(a)-1)^2$.
\end{proposition}

\begin{proof}
Let us define the series $p_i = \PP(|w_{..i}|_a \le |w_{..i}|_b)$, $q_i = \PP(|w_{..i}|_a = |w_{..i}|_b)$, and $r_i = \PP(|w_{..i}|_a = |w_{..i}|_b + 1)$, giving the probabilities that $a$ is not more frequent than $b$, $a$ occurs as often as $b$, and $a$ occurs once more than $b$ among the first $i$ letters in $w$, respectively.
By definition we have, for all $i \ge 0$:
\begin{align*}
p_{2i} &= \sum_{j=0}^{i} \binom{2i}{j} (1-p(a))^{2i-j} p(a)^{j}; &
p_{2i+1} &= \sum_{j=0}^{i} \binom{2i+1}{j} (1-p(a))^{2i+1-j} p(a)^{j}; \\
q_{2i} &= \binom{2i}{i} (1-p(a))^{i} p(a)^{i}; &
q_{2i+1} &= 0; \\
r_{2i} &= 0; &
r_{2i+1} &= \binom{2i+1}{i+1} (1-p(a))^{i} p(a)^{i+1}.
\end{align*}
Furthermore, observe that
\begin{align}\label{eq:recurrence}
  p_{i+1} = p_{i} + (1-p(a))r_{i} - p(a) q_{i} \,.
\end{align}

We show that for each $i > 0$, $p_{2i}$ is dominated by a partial sum of the geometric series with initial value $q_{2i}$ and rate $0 \le \frac{1-p(a)}{p(a)} < 1$:
\begin{align*}
p_{2i} &= \sum_{j=0}^{i} \binom{2i}{j} (1-p(a))^{2i-j} p(a)^{j} \\
&\le \binom{2i}{i} \sum_{k=0}^{i} (1-p(a))^{i+k} p(a)^{i-k} \\
&= \binom{2i}{i} (1-p(a))^{i} p(a)^{i} \sum_{k=0}^{i} \left(\frac{1-p(a)}{p(a)}\right)^k \\
&\leq \binom{2i}{i} (1-p(a))^{i} p(a)^{i} \sum_{k=0}^{\infty} \left(\frac{1-p(a)}{p(a)}\right)^k \\
&= q_{2i} \frac{1}{1-\frac{1-p(a)}{p(a)}} \\
&= q_{2i} \frac{p(a)}{2p(a)-1} \,.
\end{align*}
Thus
\begin{equation}\label{eq:one}
q_{2i} \ge \frac{2p(a)-1}\sigma p_{2i} \,.
\end{equation}
We also have $r_{2i+1} = \frac {2i+1} {i+1} p(a) q_{2i}$, which gives us
\begin{equation}\label{eq:two}
r_{2i+1} \le 2p(a) q_{2i} \,.
\end{equation}
Now we show that $p_{2i}$ decreases at a constant rate:
\begin{align*}
p_{2i+2} &= p_{2i} + (1-p(a))r_{2i+1} - p(a) q_{2i} && \text{by~\eqref{eq:recurrence}} \\
&\le p_{2i} + 2(1-p(a))p(a) q_{2i} - p(a) q_{2i} && \text{by~\eqref{eq:two}} \\
&= p_{2i} + p(a) q_{2i} - 2p(a)^2 q_{2i} && \\
&= p_{2i} - (2p(a)-1)p(a) q_{2i} && \\
&\le p_{2i} - (2p(a)-1)^2 p_{2i} && \text{by~\eqref{eq:one}} \\
& = (1-(2p(a)-1)^2)p_{2i} \,. && 
\end{align*}
Let $\rho = 1-(2\sigma-1)^2$. Since $p_0 = 1$ and $0 \le \rho < 1$, for all $i \ge 0$ we get
\begin{equation}\label{eq:three}
p_{2i} \le \rho^i \,,
\end{equation}
and knowing that $r_{2i} = 0$, we get
\begin{equation}\label{eq:four}
p_{2i+1} \le p_{2i} \,,
\end{equation}
which concludes our proof.
\end{proof}


\end{document}